\documentclass[12pt,a4paper,bezier]{article}
\usepackage{graphics,epsfig,makeidx}
      
\newfont{\mathea}{msam10 scaled\magstep0}
\newfont{\matheb}{msbm10 scaled 1095}
\newfont{\tmpEins}{cmsy10 scaled 2074}
\newfont{\tmpZwei}{cmsy10 scaled 1095}
\newfont{\tmpDrei}{cmsy10 scaled 1000}
\newfont{\tmpVier}{cmsy5 scaled 1000}
\newfont{\tmpFuenf}{msbm7 scaled\magstep0}

\def\dach#1#2{\mbox{$\mathop{\vbox{\ialign{%
  $##\crcr\hfil #1 \hfil$\crcr}}}\limits^{\scriptscriptstyle #2}$}}

\def\rnzs{\dach{\rho_2}{\mbox{$\scriptscriptstyle\kern-.7mm0$}}\kern-1.2mm'}

\def\Subset{\mbox{$\subset\kern-.5mm\subset$}}
\newcommand{\LI}{\mbox{{\rm L$^{\kern-.15em\raise.2ex\hbox{\scriptsize 1}}$}}}

\def\Ldummy{\left.\bgroup}
\def\Rdummy{\egroup^{\rule{0mm}{1.4mm}}\right.}
\def\LA{\left\langle\bgroup}
\def\RA{\egroup^{\rule{0mm}{1.4mm}}\right\rangle_{\cal A}^{}}
\def\LR{\left(\bgroup}
\def\RR{\egroup^{\rule{0mm}{1.4mm}}\right)}
\def\LG{\left\{\bgroup}
\def\RG{\egroup^{\rule{0mm}{1.4mm}}\right\}}

\def\Wort#1{\mbox{{\rm #1\kern.1em}}}

\def\lfac#1#2{\vcenter{\hbox{$#1\kern-.2em\raise-.6ex\hbox{\Large{/}}%
 \kern-.2em\raise-1.2ex\hbox{$#2$}$}}}

\def\gin{\mbox{\tmpZwei\symbol{91}\kern-1.4mm\rule{.2mm}{1.85mm}\kern1.4mm}}
\def\gni{\mbox{\tmpZwei\symbol{92}\kern-1.4mm\rule[.15mm]{.2mm}{1.85mm}%
  \kern1.4mm}}

\def\EINS{{\mathop{1\kern-.25em\mbox{{\rm{\small l}}}}}}

\begin{document}
{\Large
\begin{center}
On the border lines between the regions of distinct solution type for solutions of the Friedmann equation
\end{center}
{\normalsize

\begin{center}
Hellmut Baumg\"artel\\
Mathematical Institute\\
University of Potsdam\\
Germany\\
e-mail: baumg@uni-potsdam.de
\end{center}

\begin{abstract}
It is well-known that there are four distinct basic types (two Big Bang types, Lemaitre and Big Crunch type) for solutions of the general Friedmann equation with positive cosmological constant, where radiation and matter do not couple (see e.g. [2, p.7]). In that paper the system of  case distinction parameters contains a "critical radiation parameter" $\sigma_{cr}$. The present note contains the constructive description of the so-called {\em border lines} between Big Bang/Big Crunch type and Big Bang/Lemaitre type for so-called Hubble solutions of the Friedmann equation by two smooth function branches, expressing the cosmological constant as unique functions of the matter and radiation density (which is considered as a parameter). These functions satisfy simple asymptotic relations w.r.t. the matter density. They are constructed as the solutions of the equation $\sigma=\sigma_{cr}$.

\end{abstract}

\section{Introduction}

Einstein's famous field equation of the general relativity theory, linking together the gravitational field with the stress-energy tensor, is also a basic tool in cosmology to solve problems of the large-scale structure of the cosmos. The idea of the existence of a definite large-scale structure of the cosmos is expressed by the "cosmological principle" (E.A. Milne 1933). This principle appears as a boundary condition of Einsteins field equation. This boundary condition consists of a special ansatz of the gravitational field given by a so-called Robertson-Walker(RW-)metric, and of the special form of of the stress-energy tensor as that of a perfect fluid (Wald [1,p.69]). Moreover, it is assumed that matter and radiation do not couple. Then Einstein's field equation reduces to the so-called Friedmann equation
\begin{equation}
\left(\frac{dR}{dt}\right)^{2}=\frac{\alpha}{R}+\frac{\sigma}{R^{2}}+\frac{1}{3}\Lambda c^{2} R^{2}-\epsilon c^{2}
\end{equation}
for the so-called scale-factor $R(t)>0$, a dimensionless number. The constant parameters are: velocity of light $c$, cosmological constant $\Lambda$, curvature constant $\epsilon=0,\pm 1$, matter invariant 
\begin{equation}
\alpha:=\frac{8\pi G}{3}\rho_{mat}(\tau)R(\tau)^{3},
\end{equation}
and radiation invariant
\begin{equation}
\sigma:=\frac{8\pi G}{3}\rho_{rad}(\tau)R(\tau)^{4},
\end{equation}
where $\rho_{mat}$ and $\rho_{rad}$ denote mass and radiation density. Note that the time invariance of the right hand sides of (2) and (3) is an implication of the assumption that matter and radiation do not couple.

Admissible initial conditions $t_{0},R_{0}$ for a solution of equ. (1) have the property that the right hand side of (1) is positive for
 $R:=R_{0}$. For an admissible solution $R(\cdot)$ the so-called Hubble parameter $t\rightarrow H(t)$ is defined by
\begin{equation}
H(t):=\left(\frac{1}{R}\frac{dR}{dt}\right)(t).
\end{equation} 
Solutions of (1) whose Hubble parameter at the present time T coincides with the famous {\em Hubble constant} $H_{0},\; H(T)=H_{0}$, are called {\em Hubble solutions} in this note.

Using the equations (2),(3), the Hubble solutions satisfy the equation
\begin{equation}
H_{0}^{2}=\frac{8\pi G}{3}\rho_{mat}(T)+\frac{8\pi G}{3}\rho_{rad}(T)+\frac{1}{3}\Lambda c^{2}-\epsilon\frac{c^{2}}{R(T)^{2}}.
\end{equation}
In the special case $\Lambda=0,\sigma=0,\epsilon=0$ it reads
\begin{equation}
H_{0}^{2}=\frac{8\pi G}{3}\rho_{mat}(T).
\end{equation}
This case is the Euclidean Einstein-de Sitter model of 1932. Since that time the term
\[
\rho_{cr}:=\frac{3}{8\pi G}H_{0}^{2}
\]
is called the {\em critical density}. Introducing instead of
$\rho_{mat}(T),\rho_{rad}(T),\Lambda$ the variables
\begin{equation}
x:=\frac{\rho_{mat}(T)}{\rho_{cr}},\quad y:=\frac{\frac{1}{3}\Lambda c^{2}}{H_{0}^{2}},
\quad z:=\frac{\rho_{rad}(T)}{\rho_{cr}},
\end{equation}
then equ. (5) can be written as
\begin{equation}
1=x+y+z-\epsilon\frac{c^{2}}{H_{0}^{2}R(T)^{2}}
\end{equation}
and the constants $\alpha,\sigma,\Lambda$ are given by
\begin{equation}
\alpha=xH_{0}^{2}R(T)^{3},\;\sigma=zH_{0}^{2}R(T)^{4},\;\Lambda=y\frac{3H_{0}^{2}}{c^{2}}.
\end{equation}
If $x+y+z-1\neq 0$ then
\begin{equation}
R(T)^{2}=\left(\frac{c}{H_{0}}\right)^{2}\frac{\epsilon}{x+y+z-1},
\end{equation}
i.e. in this case the scale-factor $R(T)$ in the present time is uniquely determined. Moreover one obtains:
\[
\mbox{If}\; x+y+z>1\; \mbox{then}\;\epsilon=+1,\quad
\mbox{if}\; x+y+z<1\; \mbox{then}\;\epsilon=-1.
\]
That is,
\begin{equation}
x+y+z=1\;\mbox{iff}\;\epsilon=0.
\end{equation}
This means: the Euclidean case is quite singular, realized only on the plane (11) and in this case the scale-factor is {\em not} uniquely determined. In any case, the geometric structure of the cosmos is uniquely determined by the expression $x+y+z-1$. Sometimes the term $\frac{c}{H_{0}}$ is called the Hubble radius.

\section{Discriminant and critical radiation parameter}

The right hand side of equation (1) can be written in the form
\[
\frac{\Lambda c^{2}}{3R^{2}}p(R),
\]
where
\[
p(R):= Rq(R)+\frac{3}{\Lambda c^{2}}\sigma,
\]
and
\[
q(R):=R^{3}-\frac{3\epsilon}{\Lambda}R+\frac{3\alpha}{\Lambda c^{2}}.
\]
is a polynomial of the third degree with discriminant
\[
\Delta:=\frac{1}{4\Lambda^{3}c^{4}}\left(9\alpha^{2}\Lambda-4\epsilon c^{4}\right).
\]
Obviously, the type of a solution of (1) depends on the mutual position of $R(T)$ and the positive zeros of $p$. The shape of $p$ is independent of $\sigma$. This fact suggests to use the polynomial $q$ and its discriminant as a first parameter to distinguish different types of solutions.

If $R(\cdot)$ is a Hubble solution then $\Delta$ can be written in the form
\[
\Delta=\frac{1}{4\Lambda^{3}}\left(\frac{R(T)H_{0}}{c}\right)^{6}D,
\]
where
\begin{equation}
D=D(x,y;z):= 27x^{2}y-4(x+y+z-1)^{3},\quad x\geq 0,y\geq 0, z\geq 0.
\end{equation}

Note that for $z=1$ one has
\[
D(x,y;1)= 27x^{2}y-4(x+y)^{3}=-(x-2y)^{2}(4x+y)\leq 0,
\]
hence $D(x,y;1)=0$ if and only if $x=2y$.

For $z>1$ one has
\[
-D(x,y;z)=4(x+y+z-1)^{3}-27x^{2}y > 4(x+y)^{3}-27x^{2}y = -D(x,y;1)\geq 0,
\]
and
$D(x,y;z)<D(x,y;1)\leq 0$, i.e. $D(x,y;z)<0$ for all $z>1$.

This means: only in the case $0\leq z<1$ there are regions $D>0$  in the first quadrant $x\geq 0,y\geq 0$.

If $D(x,y;z)<0$ then we introduce the angle $\phi,\frac{\pi}{2}<\phi<\pi$ by
\begin{equation}
\cos\phi=-\frac{\sqrt{27}}{2}\frac{xy^{\frac{1}{2}}}{(x+y+z-1)^\frac{3}{2}},\quad
\frac{\pi}{2}<\phi<\pi,
\end{equation}
which is associated to the triple $(x,y;z)$. The boundary value $\phi=\pi$ characterizes the triples $(x,y;z)$, where $D=0$, i.e. the boundary of the region $D<0$, for example in the case $z=1$ the line $2x=y$. That is: fixing a parameter $z=z_{0}$, then by equation (13) to each $(x,y)$ of the first quadrant $x>0,y>0$ with $D<0$ is associated a uniquely
determined angle $\phi,\frac{\pi}{2}<\phi<\pi$ and the algebraic curves $\phi=const$ exhaust the region $D<0$ within the first quadrant $x>0,y>0$. In the following these curves are called the $\phi$-curves.

The second parameter to distinguish different types of solutions of (1) is the critical radiation parameter $\sigma_{cr}$, introduced in [2], defined for triples, where $D<0$, by
\begin{equation}
\sigma_{cr}=\frac{c^{4}}{H_{0}^{2}}\frac{2}{3}y^{-1}\left(\cos\frac{\psi}{3}\right)^{2}
\left(2\left(\cos\frac{\psi}{3}\right)^{2}-1\right),\quad \cos\psi=\frac{1}{\sqrt{2}}\cos\phi,
\quad \frac{3\pi}{4}>\psi>\frac{\pi}{2}
\end{equation}
a slight modification compared with the expression (12) in [2]. It should be emphasized that $\sigma_{cr}$ is - according to (12) and (13) - a function of $x,y;z$, just as $\sigma$ is according to (9) and (10).

$D$ and $\sigma_{cr}$ are sufficient to describe the two essential statements on the  zeros of the polynomial $p$ (see [2, Sec. III)
\begin{itemize}
\item[(i)]
If $D<0$ and $0\leq\sigma<\sigma_{cr}$ then there are two positive zeros $R_{1}<R_{2}$ (if $x=0$ then $R_{1}=0$).
\item[(ii)]
If $D\geq0$ and $\sigma>0$ or $D<0$ and $\sigma>\sigma_{cr}$ then there is no positive zero.
\end{itemize}

Hence in the case $D<0$ the $\sigma$-value $\sigma:=\sigma_{cr}$ separates the cases (i) and (ii), i.e. the solutions of the equation $\sigma=\sigma_{cr}$ are the corresponding border lines within the region $D<0$.

In the case $D<0$ and $\sigma=0$, i.e. $z=0$, one obtains the explicit expressions
\begin{equation}
R_{1}(x,y;0):=\frac{c}{H_{0}}\frac{2}{\sqrt{3}}y^{-\frac{1}{2}}\cos\left(\frac{1}{3}(2\pi-
\phi)\right),\quad R_{2}(x,y;0):=\frac{c}{H_{0}}\frac{2}{\sqrt{3}}y^{-\frac{1}{2}}
\cos\frac{1}{3}\phi.
\end{equation}

Since in the case (i) one has $p(R)<0$ for $R\in(R_{1},R_{2})$, this is an forbidden interval for initial conditions $t,R$. However, for Hubble solutions, $R(T)=R(x,y;z)$ given by equation (10) is admissible, i.e. one obtains that either
\begin{equation}
R(x,y;z)>R_{2}(x,y;z)
\end{equation}
or
\begin{equation}
R(x,y;z)<R_{1}(x,y;z).
\end{equation}

\section{The equation $\sigma=\sigma_{cr}$}

Also in the following $z$ is considered as a 
parameter and the equation is considered in the first quadrant $x\geq 0,y\geq 0$. As already mentioned,
the terms $\sigma$ and $\sigma_{cr}$ are functions of $(x,y)$ and of the parameter $z$.
The function $\sigma_{cr}$ is given by (13). According to equations (9) and (10) one obtains for $\sigma$
\begin{equation}
\sigma=\frac{c^{4}}{H_{0}^{2}}\frac{z}{(x+y+z-1)^{2}}.
\end{equation}
Then the equation $\sigma=\sigma_{cr}$ reads
\begin{equation}
\frac{4zy}{(x+y+z-1)^{2}}=\frac{8}{3}\left(\cos\frac{\psi}{3}\right)^{2}
\left(2\left(\cos\frac{\psi}{3}\right)^{2}-1\right),
\end{equation}
where $\psi$ is a function of $x,y$ and $z$, according to equ. (13) and (14). Solution of equation (19) means the construction of functions
$x\rightarrow Y(x;z_{0}),x\geq 0$, for every parameter $z_{0}>0$, which satisfy this equation. As a function of $\phi$, the right hand side of equ. (19) has a simple structure. In the following we put
\begin{equation}
\frac{8}{3}\left(\cos\frac{\psi}{3}\right)^{2}\left(2\left(\cos\frac{\psi}{3}\right)^{2}-
1\right)=:F(\phi),\quad \frac{\pi}{2}\leq\phi\leq\pi.
\end{equation}
One obtains $F(\frac{\pi}{2})=1,F(\pi)=0$ and $F(\cdot)$ is strongly monotonically decreasing.

\vspace{3mm}

The proofs for the solution of equ. (19) are considered separately for the parameter regions $0\leq z<1,z=1$ and $z>1$. The case $z=1$ is considered first. On the one hand, it is a rather singular case, but on the other hand it is explicitly soluble.

\section{Results}
\subsection{Theorem}

(i)Let $z=1$. Then
the solution of equ. (19) consists of two functions (branches)
\[
(0,\infty)\ni x\rightarrow Y_{\pm}(x),
\]
given by the parameter representation
\[
x:=x_{\pm}(\phi),\quad Y_{\pm}(x):=\mu_{\pm}(\phi)x_{\pm}(\phi),\quad \frac{\pi}{2}<
\phi<\pi,
\]
where
\begin{equation}
\mu_{\pm}(\phi):=3\nu_{\pm}(\phi)-1,\quad \nu_{+}(\phi):=\frac{\cos\frac{1}{3}\phi}
{\mid\cos\phi\mid},\quad \nu_{-}(\phi):=\frac{\cos\frac{1}{3}(2\pi-\phi)}
{\mid\cos\phi\mid}
\end{equation}
and
\begin{equation}
x_{\pm}(\phi):=\frac{4}{9}\frac{1}{F(\phi)}\left(\frac{3}{\nu_{\pm}(\phi)}-
\frac{1}{\nu_{\pm}(\phi)^{2}}\right).
\end{equation}
The solutions $Y_{\pm}(\cdot)$ have the properties
\begin{equation}
0<Y_{-}(x)<\frac{1}{2}x<Y_{+}(x),\quad x>0,
\end{equation}
such that
\[
Y_{\pm}(x)=\frac{1}{2}x \pm u_{\pm}(x),\quad u_{\pm}(x)>0,
\]
\begin{equation}
\lim_{x\rightarrow 0}Y_{+}(x)=4,\quad \lim_{x\rightarrow 0}Y_{-}(x)=0,
\end{equation}
and
\begin{equation}
\lim_{x\rightarrow\infty}\frac{u_{\pm}(x)}{x}=0.
\end{equation}


\hspace{1cm}\includegraphics[width=0.7\textwidth]{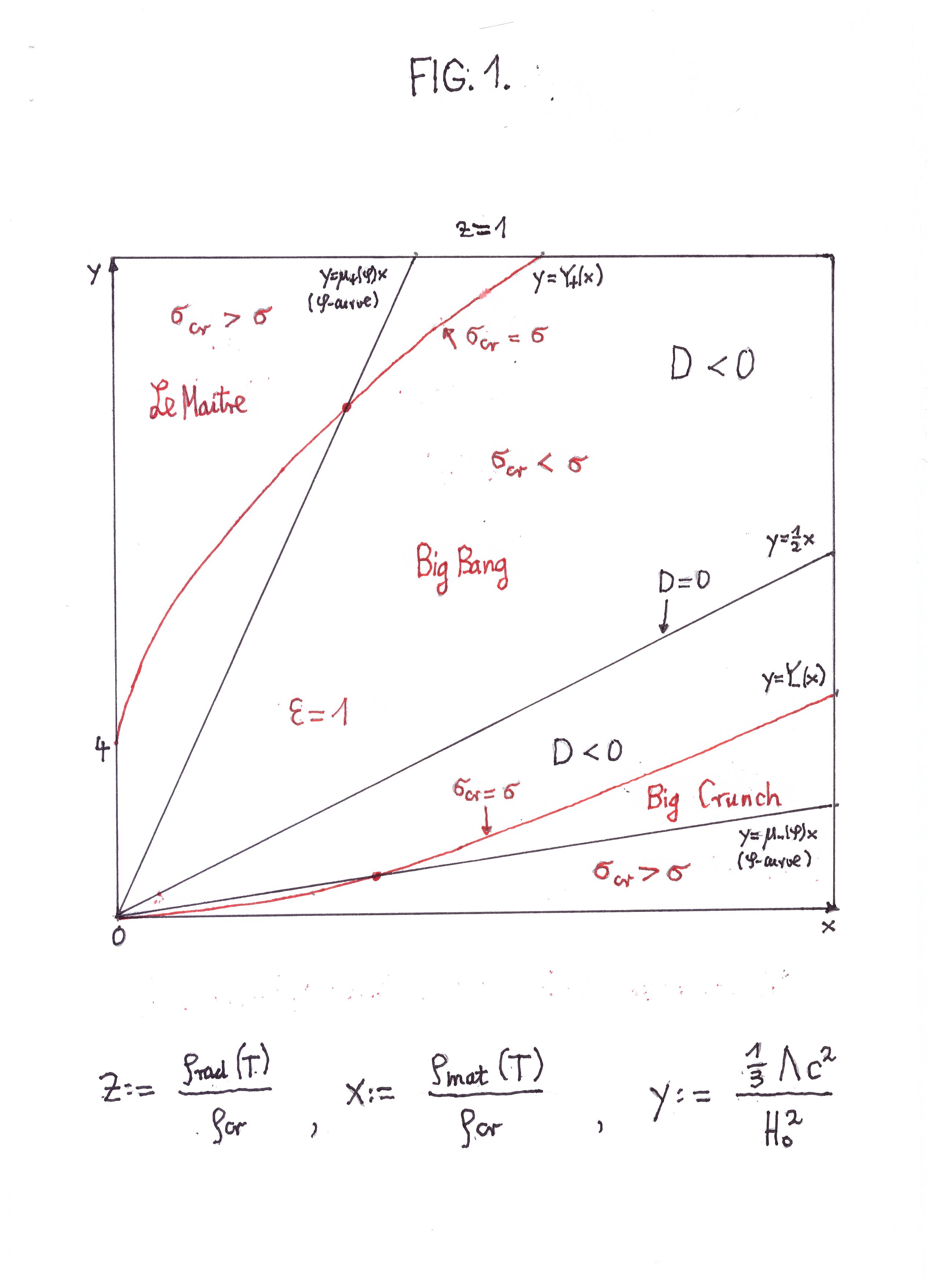}

                             
(ii)Let $z>1$. Then the solution of equ. (19) consists of two functions (branches)
\[
(0,\infty)\ni x\rightarrow Y_{\pm}(x)
\]
given by the parameter representation
\begin{equation}
x=x(\phi,\rho_{\pm}(\phi)),\quad Y_{\pm}(x)=\rho_{\pm}(\phi)x(\phi,\rho_{\pm}(\phi)),\quad
\frac{\pi}{2}<\phi<\pi,
\end{equation}
where
\begin{equation}
x(\phi,\mu):=(z-1)\frac{2^{2/3}(\cos\phi)^{2/3}}{3\mu^{1/3}-2^{2/3}(\cos\phi)^{2/3}(\mu+1)},
\quad \mu_{-}(\phi)<\mu<\mu_{+}(\phi),
\end{equation}
and the corresponding $\phi$-curve is given by $y_{\pm}(x,\phi):=\mu x(\phi,\mu)$.
The terms $\mu:=\rho_{\pm}(\phi)$ are uniquely determined solutions of the equation
\begin{equation}
(\cos\phi)^{2/3}\mu^{1/3}(3\mu^{1/3}-2^{2/3}(\cos\phi)^{2/3}(\mu+1))=
\frac{3^{2}}{2^{8/3}}\frac{z-1}{z}F(\phi),
\end{equation}
where
\begin{equation}
\mu_{-}(\phi)<\rho_{-}(\phi)<\rho_{+}(\phi)<\mu_{+}(\phi).
\end{equation}
The solutions $Y_{\pm}(\cdot)$ have the properties $0<Y_{-}(x)<Y_{+}(x),\;x>0$,
\begin{equation}
0<y_{-}(x;\phi)<Y_{-}(x)<\frac{1}{2}x<Y_{+}(x)<y_{+}(x;\phi),\quad x\;\mbox{sufficiently large}, \quad \frac{\pi}{2}<\phi<\pi,
\end{equation}
such that
\[
Y_{\pm}(x)=\frac{1}{2}x\pm u_{\pm}(x),\quad u_{\pm}(x)>0, \quad x\;\mbox{sufficiently large},
\]
\begin{equation}
\lim_{x\rightarrow 0}Y_{\pm}(x)=(1\pm\sqrt{z})^{2},
\end{equation}
\begin{equation}
\lim_{x\rightarrow\infty}\frac{u_{\pm}(x)}{x}=0.
\end{equation}


\includegraphics[width=0.7\textwidth]{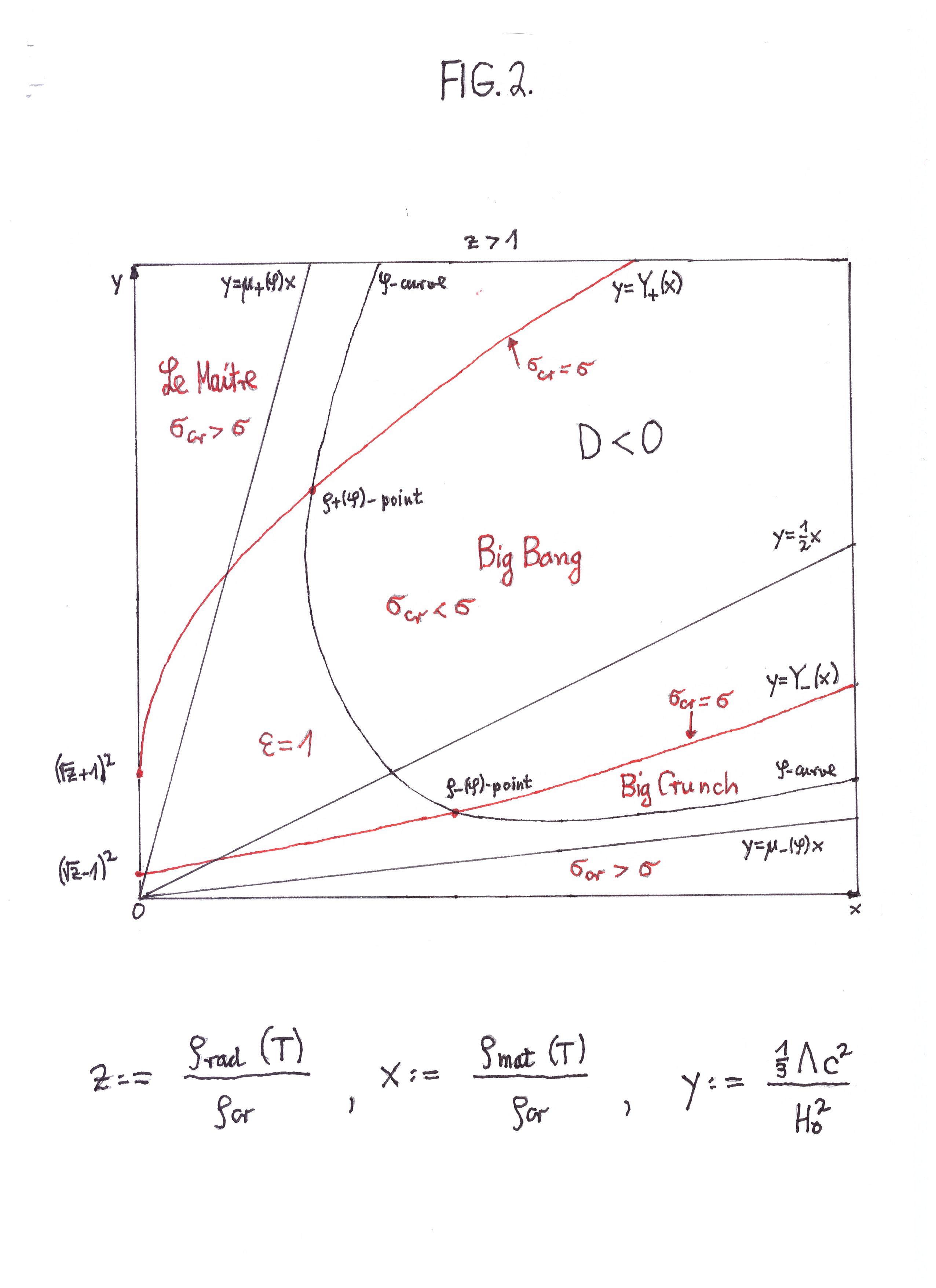}


(iii) Let $0<z<1$.
In this case every $\phi$-curve, $\frac{\pi}{2}<\phi<\pi$ has two branches, a lower one $x\rightarrow y_{-}(x;\phi)$ for $x\geq a$, where $a:=1-z$, starting at
$(a,0)$, and an upper one $x\rightarrow y_{+}(x;\phi)$ for $x\geq 0$, starting at $(0,a)$.
The branches $y_{\pm}(\cdot;\pi)$ of the $\pi$-curve form the boundary of the region defined by $D<0$. It corresponds to $D=0$ (cf. Sec. 2).
The solution of equ. (19) consists of two functions (branches)
\[
(0,\infty)\ni x\rightarrow Y_{+}(x),\quad (a,\infty)\ni x\rightarrow Y_{-}(x).
\]
The branch $Y_{+}(\cdot)$ is given by the parameter representation
\begin{equation}
x=x(\phi,\rho_{+}(\phi)),\quad Y_{+}(x)=\rho_{+}(\phi)x(\phi,\rho_{+}(\phi)),\quad\frac{\pi}{2}<\phi<\pi,
\end{equation}
where
\begin{equation}
x(\phi,\mu):=(1-z)\frac{2^{2/3}(\cos\phi)^{2/3}}{2^{2/3}(\cos\phi)^{2/3}(\mu+1)-3\mu^{1/3}},\quad \mu>\mu_{+}(\phi),
\end{equation}
and the term $\mu:=\rho_{+}(\phi)$ is the uniquely determined solution of the equation
\begin{equation}
(\cos\phi)^{2/3}\mu^{1/3}(2^{2/3}(\cos\phi)^{2/3}(\mu+1)-3\mu^{1/3})=
\frac{3^{2}}{2^{8/3}}\frac{1-z}{z}F(\phi),
\end{equation}
where $\rho_{+}(\phi)>\mu_{+}(\phi)$. This means: every $\phi$-curve has exactly one intersection with the branch $Y_{+}(\cdot)$, realized by the upper branch $y_{+}$ of the $\phi$-curve.

In contrast to this property of $Y_{+}(\cdot)$ the branch $Y_{-}(\cdot)$ has either exactly two intersections with a $\phi$-curve or there is no intersection. More precisely:  

To every $a,0<a<1$, there is an angle $\phi(a),\frac{\pi}{2}<\phi(a)<\pi$, such that the parameter representation of $Y_{-}(\cdot)$ is given by
\begin{equation}
x=x(\phi,\rho^{\pm}_{-}(\phi)),\quad Y_{-}(x)=\rho^{\pm}_{-}(\phi)x(\phi,\rho^{\pm}_{-}(\phi)),\quad \phi(a)<\phi<\pi,
\end{equation}
where
\begin{equation}
0<\rho^{-}_{-}(\phi)<\mu_{max}<\rho^{+}_{-}(\phi)<\mu_{-}(\phi),
\end{equation}
and the left hand side of equ. (35) takes its maximum at $\mu_{max}$. Note that for $\phi:=\phi(a)$ the terms $\rho^{-}_{-}(\phi(a))$ and $\rho^{+}_{-}(\phi(a))$coincide, $\rho_{-}^{+}(\phi(a))=\rho_{-}^{-}(\phi(a))$. If $\phi<\phi(a)$ then there is no intersection with this $\phi$-curve.

The solutions $Y_{\pm}(\cdot)$ have the properties
\[
0<Y_{-}(x)<\frac{1}{2}x<Y_{+}(x),\; x>a,\quad 0<Y_{+}(x),\;x>0
\]
\begin{equation}
0<y_{-}(x,\phi)<Y_{-}(x)<\frac{1}{2}x<Y_{+}(x;)<y_{+}(x,\phi),\quad x\;\mbox{sufficiently large},
\end{equation}
\[
Y_{\pm}(x)=\frac{1}{2}x\pm u_{\pm}(x),\quad u_{\pm}(x)>0,
\]
\begin{equation}
\lim_{x\rightarrow a}Y_{-}(x)=0,
\end{equation}
\begin{equation}
\lim_{x\rightarrow 0}Y_{+}(x)=(1+\sqrt{z})^{2},
\end{equation}
\begin{equation}
\lim_{x\rightarrow\infty}\frac{u_{\pm}(x)}{x}=0.
\end{equation}


\includegraphics[width=0.7\textwidth]{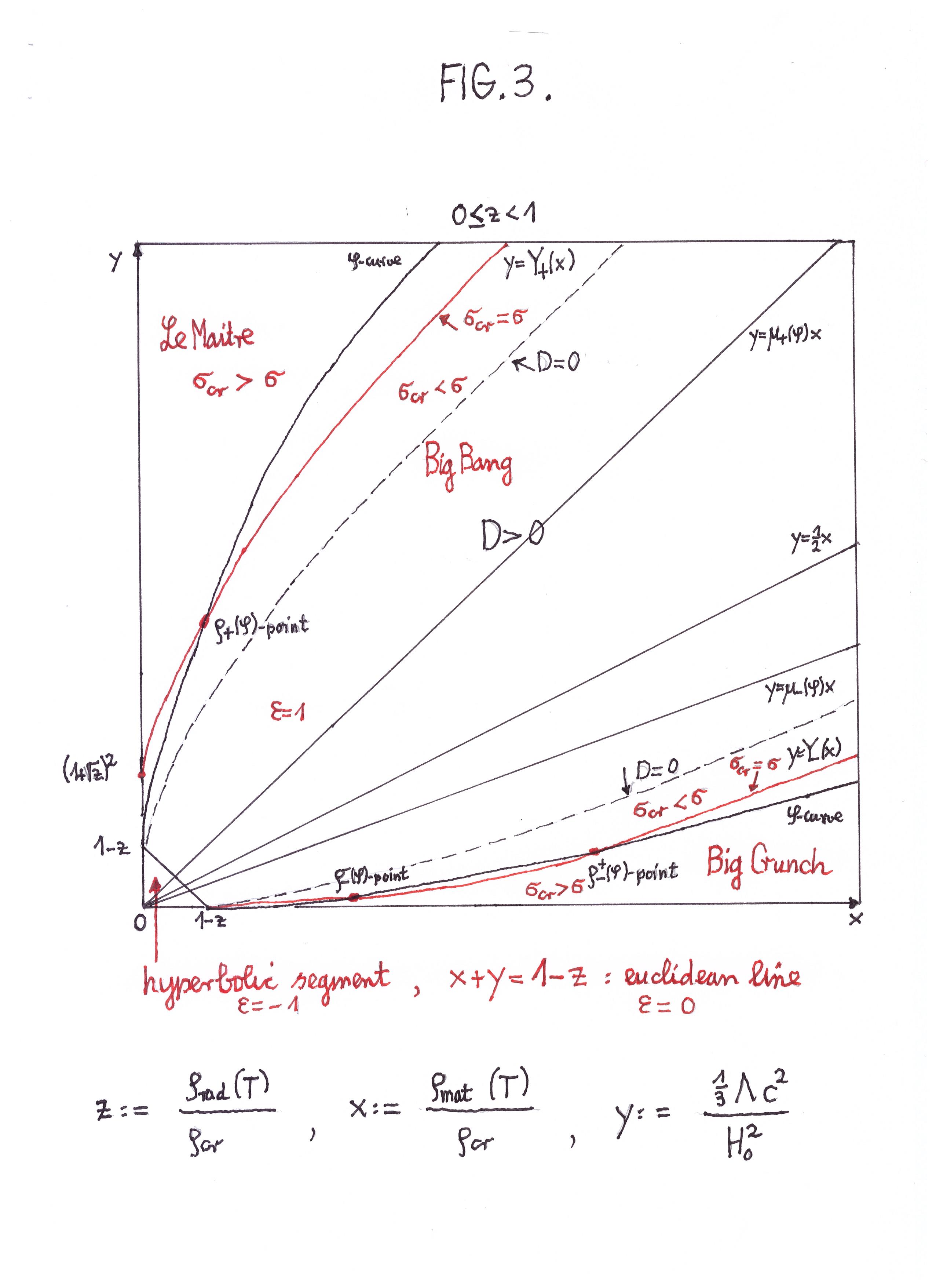}


(iv) Let $z=0$. In this case one has $\sigma=0$, i.e. the solution branches of the equation $\sigma=\sigma_{cr}$ coincide with the branches $y_{\pm}(\cdot;\pi)$ of the $\pi$-curve
\[
4(x+y-1)^{3}=27x^{2}y,
\]
i.e. they coincide with the curve where $D=0$.

\subsection{The regions $G_{\pm}(z)$}

In the case $D(x,y;z)<0$ these regions are defined for $z\geq 1$ by
\[
G_{+}(z):=\{(x,y):x>0,y>Y_{+}(x;z)\},\quad G_{-}(z):=\{(x,y):x>0,y<Y_{-}(x;z)\},
\]
and for $0\leq z<1$ by
\[
G_{+}(z):=\{(x,y):x>0,y>Y_{+}(x:z)\},\quad G_{-}(z):=\{(x,y):x>1-z,y<Y_{-}(x;z)\},
\]
such that the region $D(x,y;z)<0$ is the union $G_{+}(z)\cup G_{-}(z)$.

These regions can be also characterized by the mutual position of $R(T)=R(x,y;z)$ and the roots $R_{1}(x,y;z)<R_{2}(x,y;z)$ of the polynomial $p(\cdot)$. Actually one obtains

\subsection{Lemma}

The regions $G_{\pm}$ can be characterized as follows:
\begin{equation}
G_{+}(z)=\{(x,y):R(x,y;z)>R_{2}(x,y;z)\}\quad G_{-}(z):=\{(x,y):R(x,y;z)<R_{1}(x,y;z)\}.
\end{equation}

\section{Proofs}
\subsection{Proof of the Theorem}
(i) Let $z=1$.
First note that
\[
\mu_{+}(\phi)\rightarrow\infty\quad\mbox{and}\quad\mu_{-}(\phi)\rightarrow 0\quad
\mbox{for}\quad\phi\rightarrow\frac{\pi}{2},
\]
further
\begin{equation}
\mu_{-}(\phi)<\frac{1}{2}<\mu_{+}(\phi),\;\frac{\pi}{2}<\phi<\pi;\quad\mu_{\pm}(\phi)
\rightarrow\mu_{\pm}(\pi)=\frac{1}{2}\quad\mbox{for}\;\phi\rightarrow\pi.
\end{equation}
The equation for the $\phi$-curves reads
\[
4(x+y)^{3}(\cos\phi)^{2}=27x^{2}y,\quad \frac{\pi}{2}<\phi<\pi.
\]
The terms $\nu_{\pm}(\phi)$ satisfy the relation
\[
4\nu_{\pm}(\phi)^{3}(\cos\phi)^{2}+1=3\nu_{\pm}(\phi)
\]
and for the terms $\mu_{\pm}(\phi)$ one obtains
\begin{equation}
27\mu_{\pm}(\phi)-4(1+\mu_{\pm}(\phi))^{3}(\cos\phi)^{2}=0.
\end{equation}
In the case $\phi=\pi$ for the left hand side of (44) one has $4(1+\mu)^{3}-27\mu=(2\mu-1)^{2}(\mu+4)$. Moreover one obtains
\begin{equation}
27\mu-4(1+\mu)^{3}(\cos\phi)^{2}>0,\quad \mu_{-}(\phi)<\mu<\mu_{+}(\phi).
\end{equation}
Therefore, the $\phi$-curve consists of the two rays
\begin{equation}
y=\mu_{\pm}(\phi)x,\;x>0,\quad \frac{\pi}{2}<\phi<\pi.
\end{equation}
In the case $\phi=\pi$ there is only one ray and for $\phi=\frac{\pi}{2}$ the limit rays are the half axes $x=0$ and $y=0$.

The equation $\sigma=\sigma_{cr}$ reads
\[
4y=(x+y)^{2}F(\phi).
\]
For a point $(x,y)$ on a $\phi$-ray (46) this means
\[
4\mu_{\pm}(\phi)x=F(\phi)(1+\mu_{\pm}(\phi))^{2}x^{2}.
\]
The solution of this equation is given by (22) which implies 
$x_{-}(\phi)<x_{+}(\phi)$ and
\[
x_{\pm}(\phi)\rightarrow 0\quad \mbox{if}\;\phi\rightarrow\frac{\pi}{2},\quad
x_{\pm}(\phi)\rightarrow\infty\quad \mbox{if}\;\phi\rightarrow\pi.
\]
Then $\mu_{-}(\phi)x_{-}(\phi)\rightarrow 0$ and $\mu_{+}(\phi)x_{+}(\phi)\rightarrow 4$
for $\phi\rightarrow\frac{\pi}{2}$.
This proves (24). If $x=x_{+}(\phi_{+})=x_{-}(\phi_{-})$ then
$x\rightarrow\infty$ iff $\phi_{\pm}\rightarrow\pi$. In this case one has
 $Y_{\pm}(x)=\mu_{\pm}(\phi_{\pm})x$. Because of (43) this proves (23). Finally, if $x\rightarrow\infty$ then
$\mu_{\pm}(\phi_{\pm})\rightarrow\mu_{\pm}(\pi)=\frac{1}{2}$. This proves (25).

\vspace{3mm}

(ii) Let $z>1$.
For convenience put $a:=z-1,a>0$. The equation for the $\phi$-curves reads
\begin{equation}
4(x+y+a)^{3}(\cos\phi)^{2}=27x^{2}y,\quad\frac{\pi}{2}<\phi<\pi.
\end{equation}
Note that in this case the limit case $\phi=\pi$ is excluded because $D<0$ everywhere. The limit case $\phi=\frac{\pi}{2}$ corresponds to the half axes $x=0,y=0$. The parameter representation for the $\phi$-curve using $\mu$-rays $y=\mu x,\mu>0$ yields the term
$x=x(\phi,\mu)$ given by equ. (27). According to equations (44) and (45) one has
$x\rightarrow\infty$ for $\mu\rightarrow\mu_{\pm}(\phi)$. That is, in this case the $\phi$-curve (47) has only a single branch, where 
$x\geq a\frac{\mid\cos\phi\mid}{1-\mid\cos\phi\mid}$ for fixed $\phi$. Moreover, the expression $3\mu^{1/3}-2^{2/3}(1+\mu)(\cos\phi)^{2/3}$ takes its maximum at
$\mu:=\frac{1}{2(\cos\phi)^{2/3}}$, i.e. for any 
$x>a\frac{\mid\cos\phi\mid}{1-\mid\cos\phi\mid}$ there are parameters $\mu_{+},\mu_{-}$ such that
\[
\mu_{-}<\frac{1}{2(\cos\phi)^{2/3}}<\mu_{+}
\]
and $x=x(\phi,\mu_{-})=x(\phi,\mu_{+})$. The corresponding values for $y$ are
$y_{\pm}(x,\phi)=\mu_{\pm}x$, i.e.one obtains
\begin{equation}
\lim_{x\rightarrow\infty}\frac{y_{\pm}(x,\phi)}{x}=\mu_{\pm}(\phi).
\end{equation}

The equation (19) for a point $(x,y)=(x,\mu x)$ on a fixed $\phi$-curve leads the equation (28). The left hand side of this equation is positive for parameters $\mu$ satisfying
$\mu_{-}(\phi)<\mu<\mu_{+}(\phi)$  and vanishes for $\mu=\mu_{\pm}(\phi)$. It
takes its maximal value in this interval at
\[
\mu_{max}^{1/3}:=\frac{2^{1/6}}{\mid\cos\phi\mid^{1/3}}\cos\frac{\psi}{3},
\]
where the angle $\psi$ is given by equ. (14). The corresponding maximum is given by
$2^{1/3}\cos\frac{\psi}{3}(2(\cos\frac{\psi}{3})^{3}-3\cos\psi)$. 
Now the inequality
\[
2^{1/3}\cos\frac{\psi}{3}\left(2\left(\cos\frac{\psi}{3}\right)^{3}-3\cos\psi\right)>\frac{3^{2}}{2^{8/3}}F(\phi)
\]
is true, it is equivalent with the inequality $\frac{1}{2}\sqrt{3}>\cos\frac{\psi}{3}$ and
$\frac{1}{2}\sqrt{3}$ is the maximal value of $\cos\frac{\psi}{3}$ in the admissible interval for $\psi$ (see equ. (14)), which is taken at the limit case $\phi=\frac{\pi}{2}$.
The consequence is that for every $\phi\in (\frac{\pi}{2},\pi)$ there are exactly two solutions $\mu:=\rho_{\pm}(\phi)$ of equ. (28), where the inequality (29) is satisfied. This proves (26). The relation (30) follows from (29). The relations (31) can be obtained by solving equ. (19) directly for $x=0$ which implies $\phi=\frac{\pi}{2}$. Concerning relation (32) note that $x\rightarrow\infty$ on a fixed $\phi$-curve corresponds to $\mu\rightarrow
\mu_{\pm}(\phi)$ and for each $\phi<\pi$ there is a solution $x=x(\phi,\mu)$ with
$\mu:=\rho_{\pm}(\phi)$, according to equ. (26). Taking the limit $\phi\rightarrow\pi$ then with equ. (48) and (30) one obtains (32).

\vspace{3mm}

(iii) Let $0<z<1$.
The equation for the $\phi$-curve reads 
\begin{equation}
4(x+y-a)^{3}(\cos\phi)^{2}=27x^{2}y,\quad \frac{\pi}{2}<\phi<\pi.
\end{equation}
The limit case $\phi=\frac{\pi}{2}$ corresponds to the half axes $x=0,y\geq a$ and $y=0,
x\geq a$. The case $\phi=\pi$ corresponds to $D=0$. The parameter representation using $\mu$-rays $y=\mu x,\mu>0$, yields the term
\begin{equation}
x=x(\phi,\mu)=a\frac{2^{2/3}(\cos\phi)^{2/3}}{2^{2/3}(\cos\phi)^{2/3}(\mu+1)-3\mu^{1/3}},\quad \mu>\mu_{+}(\phi),\;\mu<\mu_{-}(\phi)
\end{equation}
for the $x$-coordinate of the point of the $\phi$-curve. The parameter values $\mu>\mu_{+}(\phi),\mu<\mu_{-}(\phi)$ describe the upper and lower branch $y_{\pm}(\cdot;\phi)$ of the $\phi$-curve, respectively. Inserting (50) with $\mu>\mu_{+}(\phi)$ and $y=\mu x$ into equ. (19) then the resulting equ. (35) is the condition for those $\mu$, where the intersection point of the $\mu$-ray with the $\phi$-curve is simultaneously a solution point of equ. (19). The left hand side of equ. (35) vanishes for $\mu:=\mu_{+}(\phi)$, it tends to infinity for $\mu\rightarrow\infty$ and it is strongly monotonically increasing for $\mu>\mu_{+}(\phi)$. If $\frac{\pi}{2}<\phi<\pi$ then $0<F(\phi)<1$ and 
$0<\frac{1-z}{z}F(\phi)<\infty$ because of $0<z<1$. That is, for every pair $\{\phi,z\}$ there is exactly one solution $\rho_{+}(\phi)$ of equ. (35) and one has $\mu_{+}(\phi)<\rho_{+}(\phi)$.

For the investigation of solutions of equ. (35) in the interval $(0,\mu_{-}(\phi))$ we write this equation in the form                   
\begin{equation}
(1-a)(\cos\phi)^{2/3}\mu^{1/3}(2^{2/3}(\cos\phi)^{2/3}(\mu+1)-3\mu^{1/3})=
\frac{3^{2}}{2^{8/3}}aF(\phi).
\end{equation}
The extrema of the left hand side of equ. (51) as a function of $\mu$ are
\[
\mu_{min}^{1/3}:=\frac{2^{1/6}}{\mid\cos\phi\mid^{1/3}}\cos\frac{\psi}{3},\quad
\mu_{max}^{1/3}:=\frac{2^{1/6}}{\mid\cos\phi\mid^{1/3}}\cos\frac{1}{3}(2\pi-\psi).
\]
First one obtains $\mu_{-}(\phi)<\mu_{min}<\mu_{+}(\phi)$. This term has been used for the solution of equ. (19) in the case $z>1$. Further, $0<\mu_{max}<\mu_{-}(\phi)$. At this term the left hand side takes its maximum. Inserting this value into the left hand side one gets
$(1-a)G(\phi)$, where
\begin{equation}
G(\phi):= 3\cdot 2^{1/3}\left(\cos\frac{1}{3}(2\pi-\psi)\right)^{2}\left(1-2\left(\cos\frac{1}{3}(2\pi-\psi)\right)^{2}\right).
\end{equation}
Recall equ. (14) which implies $\frac{\pi}{4}>\frac{\psi}{3}>\frac{\pi}{6}$ and
$\frac{5}{12}\pi<\frac{1}{3}(2\pi-\psi)<\frac{\pi}{2}$. The function
$\phi\rightarrow G(\phi),\frac{\pi}{2}<\phi<\pi$, is monotonically increasing and
\begin{equation}
G(\frac{\pi}{2})=0,\quad G(\pi)= 3\cdot 2^{1/3}\left(\cos\frac{5}{12}\pi\right)^{2}
\left(1-2\left(\cos\frac{5}{12}\pi\right)^{2}\right)>0.
\end{equation}
Comparing the maximum value $G(\phi)$ of the left hand side of equ. (51) with the right hand side $a\cdot 3^{2}2^{-8/3}F(\phi)$ and 
taking into account the monotony properties of $G(\cdot)$ and $F(\cdot)$, further (51) and $F(\frac{\pi}{2})=1,\;F(\pi)=0$ then one obtains: There is exactly one angle
$\phi(a),\;\frac{\pi}{2}<\phi(a)<\pi$, such that
\[
(1-a)G(\phi(a))=a\cdot \frac{3^{2}}{2^{8/3}}F(\phi(a)).
\]
Further, if $\frac{\pi}{2}<\phi<\phi(a)$ then the left hand side of equ. (51) at the maximum point $\mu_{max}$ is smaller than the right hand side at $\phi$; however, if $\phi(a)<\phi
<\pi$ then the left hand side of equ. (51) at the maximum point is larger than the right hand side at $\phi$. This implies: In the case $\frac{\pi}{2}<\phi<\phi(a)$ there is no solution $\mu$ of equ. (51). If $\phi(a)<\phi<\pi$ then there are exactly two solutions
$\rho_{-}^{\pm}(\phi)$ of equ. (51) such that equ. (36) is true.

From equ. (50) it follows that $\lim_{\mu\rightarrow 0}
x(\phi,\mu)=a$ for all $\phi,\frac{\pi}{2}<\phi<\pi$, i.e. the lower branch of the
$\phi$-curve starts at $(a,0)$. This implies (39), according to (37).

Relation (40) can be obtained by solving equ. (19) directly for $x=0$ which implies
$\phi=\frac{\pi}{2}$. Note that in this case the second (formal) solution
$(1-\sqrt{z})^{2}$ is excluded because of $(1-\sqrt{z})^{2}<1-z$ which implies
$D(0,(1-\sqrt{z})^{2},z)>0$, i.e. the pair $(0,(1-\sqrt{z})^{2})$ belongs to the region
$D>0$.

Concerning relation (41) note that it follows from (50) that for $\mu>\mu_{+}(\phi),\;\mu\rightarrow\mu_{+}(\phi)$ or $\mu<\mu_{-}(\phi),
\;\mu\rightarrow\mu_{-}(\phi)$ one has $x(\phi,\mu)\rightarrow\infty$. This means that for the two branches $y_{\pm}(\cdot;\phi)$ of the $\phi$-curves one has
\begin{equation}
\lim_{x\rightarrow\infty}\frac{y_{\pm}(x;\phi)}{x}=\mu_{\pm}(\phi).
\end{equation}
However, for each $\phi,\;\phi(a)<\phi<\pi$, there is a solution $x:=x(\phi,\mu)$ of equation (19), either with $\mu:=\rho_{+}(\phi)$ according to (35) or with
$\mu:=\rho_{-}^{+}(\phi)$ according to (36) and (37). Taking the limit $\phi\rightarrow\pi$ then with equ. (52) one obtains (41).

\subsection{Proof of the Lemma}

First note that, according to equations (10) and (15), in the case $z_{0}=0$ on has to show that
\[
(x+y-1)^{-1}>\frac{4}{3}y^{-1}\left(\cos\frac{\phi}{3}\right)^{2},\quad (x,y)\in G_{+},
\]
and
\[
(x+y-1)^{-1}<\frac{4}{3}y^{-1}\left(\cos\frac{1}{3}(2\pi-\phi)\right)^{2},\quad 
(x,y)\in G_{-}.
\]
These inequalities can be verified by a direct estimation.

The corresponding property is also valid for $z>0$. First there are special points satisfying this property, for example $(0,y)$, where $y>(1+\sqrt{z})^{2}$, for $G_{+}(z)$, and, on the other hand, $(x,y)$, where $x>1$ large and $0<y<1$, for $G_{-}(z)$. 
Further both regions are simply connected, i.e. an arbitrary point of a region can be smoothly joined within the region with such a special point. Choose a smooth path joining the special point $(x_{0},y_{0})$, say, of $G_{+}(z)$ with an arbitrary point $(x_{1},y_{1})$ of this region. Consider the function $R(x,y;z)-R_{2}(x,y;z)$ for points of the path. At $(x_{0},y_{0})$ it is positive. Assume that $R(x_{1},y_{1};z)<R_{1}(x_{1},y_{1};z)$, then  the value of the function at this point is smaller than 
$R_{1}(x_{1},y_{1};z)-R_{2}(x_{1},y_{1};z)<0$, i.e. then there is a point on the path such that value of the function at this point is zero which is a contradiction.              

\section{Remark}

The asymptotic relation for the functions $u_{\pm}$ points to a logarithm-like behaviour. The asymptotics of these functions could be investigated more precisely. The ray
$y=\frac{1}{2}x$ in the description of the border lines characterizes in the case of vanishing radiation the the transition points from deceleration to acceleration.

\section{References}
\begin{enumerate}
\item
R.M. Wald: General Relativity, University of Chicago Press 1984\\
\item
H. Baumg\"artel: On a critical radiation density in the Friedmann equation,\\
JMP 53, 122505 (2012)
\end{enumerate}

\end{document}